\documentclass[runningheads]{llncs}

\newcommand{\F}{\mathbb{F}}
\newcommand{\G}{\mathbb{G}}

\newcommand{\sample}{\xleftarrow{\$}}

\newcommand{\Enc}{{\sf Enc}}
\newcommand{\CheckPoint}{{\sf CheckPoint}}

\newcommand{\MapHomomorphic}{{\sf MapHomomorphic}}
\newcommand{\MapRecoverable}{{\sf MapRecoverable}}
\newcommand{\Dec}{{\sf Dec}}

\newcommand{\LSB}{{\sf LSB}}

\renewcommand{\note}{{\sf note}}

\newcommand{\PI}{{\sf PI}}
\newcommand{\sk}[1][]{{\sf sk}_{#1}} 
\newcommand{\pk}[1][]{{\sf pk}_{#1}}
\newcommand{\enc}[1][]{\textnormal{enc}_{#1}}
\newcommand{\nat}{\textnormal{nat}} 
\newcommand{\recov}{\textnormal{rec}} 
\newcommand{\homom}{\textnormal{hom}} 
\newcommand{\new}{\textnormal{new}} 
\newcommand{\old}{\textnormal{old}} 

\usepackage{geometry}
\usepackage{authblk}
\usepackage[ruled,vlined]{algorithm2e}

\usepackage[english]{babel}
\usepackage[utf8]{inputenc}
\usepackage{amsmath, mathtools}
\usepackage{amsfonts}
\PassOptionsToPackage{hyphens}{url}\usepackage{hyperref}
\usepackage{listings}
\usepackage{graphicx, color, xcolor}
\usepackage{tabularx}
\usepackage{xspace} 
\usepackage{soul} 

\usepackage[colorinlistoftodos, textwidth = 40mm]{todonotes}

\usepackage{tikz}
\usetikzlibrary{arrows, positioning, automata}	
\tikzstyle{process} = [	rectangle,
						minimum width=1cm, 
						minimum height=0.8cm,
						text centered, 
						draw=black,]
\tikzstyle{arrow} =[thick,->,>=stealth]

\newtheorem{protocol}{Protocol}

\usepackage{cite}
\usepackage[T1]{fontenc}
\usepackage{graphicx}
\usepackage{color}

\begin{document}
\title{Homomorphic Encryption-based Vaults for Anonymous Balances on VM-enabled Blockchains}
%
%
\author{Xavier Salleras}
\authorrunning{X. Salleras}
%
\institute{Dusk Network\\
\email{xavier@dusk.network}}
\maketitle              

\begin{center}
November 18, 2025
\end{center}
\begin{abstract}
    In this work, we present homomorphic encryption-based vaults (Haults), a permissioned privacy-preserving smart wallet protocol for VM-enabled blockchains that keeps users' balances confidential, as well as the amounts transacted to other parties. To comply with regulations, we include optional compliance features that allow specific entities (the auditors) to retrieve transaction amounts or execute force transfers when necessary. Our solution uses ElGamal over elliptic curves to encrypt balances, combined with zero-knowledge proofs to verify the correctness of transaction amounts and the integrity of the sender's updated balance, among other security checks. We provide a detailed explanation of the protocol, including a security discussion and benchmarks from our proof-of-concept implementation, which yield great results. Beyond in-contract issued tokens, we also provide a thorough explanation on how our solution can be compatible with external ones (e.g., Ether or any ERC20).

\end{abstract}

\section{Introduction}
\label{sec:introduction}
Blockchains have reshaped parts of the financial paradigm by enabling new infrastructures for decentralized finance (DeFi). In the particular case of Ethereum (as well as other Virtual Machine-enabled Blockchains), it allows decentralized verification of deterministic code execution through smart contracts. This enables many on-chain applications, such as issuance and management of crypto assets (e.g., ERC-20~\cite{erc20} tokens). While these systems ensure transparency and auditability, they expose sensitive financial data by design: all account balances, token transfers, and contract interactions are permanently visible on-chain. This transparency, while beneficial for protocol integrity, introduces significant privacy concerns for individuals and institutions transacting on-chain~\cite{sedlmeir2022transparency,zhou2022behavior}.

In parallel, a variety of privacy-preserving protocols have been developed over the last few years, utilizing advanced cryptographic tools such as zero-knowledge proofs, ring signatures, and stealth addresses. Projects like Zerocash~\cite{sasson2014zerocash} or Monero~\cite{van2013cryptonote} achieve strong levels of anonymity and unlinkability, offering fully confidential transactions where both identities and amounts are hidden.

While full anonymity serves many important use cases, especially in contexts of personal financial privacy, it is not always a necessary, or even desirable, requirement. In many real-world use cases, participant identities may be required to remain public for regulatory or business reasons, while only the transaction amount needs to remain confidential. Examples include institutional transfers between known counterparties, private bidding in procurement, confidential salary or bonus disbursements within decentralized autonomous organizations (DAOs), and privacy-preserving donations or grants. In such scenarios, total anonymity may complicate compliance, increase trust barriers, or invite regulatory scrutiny~\cite{belanger11privacy,rehman2021trustmodels}.

Moreover, fully private systems often come with significant technical and operational complexity. In general, they require resource-intensive setup and verification processes, heavy synchronization with encrypted ledgers, and complex scanning mechanisms to detect incoming transfers. This often hinders usability and performance in decentralized applications, especially where speed and transparency of interaction are important.

\subsection{Related Work}

There is a rich body of work on privacy-preserving transactions in blockchain systems. One of the pioneering works, Zcash~\cite{zcash2016} (based on the previous work on Zerocash~\cite{sasson2014zerocash}), introduced shielded transactions using zk-SNARKs to hide transaction details (sender, recipient, and transferred amount) while ensuring ledger integrity.

Zether~\cite{cryptoeprint:2019/191} proposed an account-based confidential transfer mechanism for Blockchains capable of deploying smart contracts, using homomorphic encryption and $\sum$-Bullets, an improvement over Bulletproofs. Like happens with Zcash, they manage to hide both identities and transaction amounts. 

Aztec~\cite{aztec} introduced new techniques for anonymous payments, including the join-split confidential transaction model. Likewise, their work on zero-knowledge proof systems like Plonk~\cite{cryptoeprint:2019/953} has contributed to the broader ecosystem of privacy-preserving protocols. As a matter of fact, Dusk~\cite{dusk} employed Plonk to implement a confidential transaction model in their own Blockchain, as well as a self-sovereign identity scheme~\cite{salleras2023citadelselfsovereignidentitiesdusk}.

On the other hand, compliance and regulatory concerns have recently increased, and some works like the protocol proposed in~\cite{BUTERIN2024100176} have addressed this issue by introducing a mechanism for users to reveal certain properties of the transactions they issue, without having to reveal their full content.

\subsection{Contributions}

In this work, we present a simple but highly practical privacy model: one where sender and recipient addresses are visible, but the transferred amount is encrypted, as well as the balances of users of such a model. We introduce \textit{homomorphic encryption-based vaults} (Haults), a privacy-preserving protocol for VM-enabled environments (e.g., EVM-compatible blockchains) that enables confidential balances and transferred amounts without compromising participant traceability. Haults are designed for semi-public financial workflows, where transparency of identities is beneficial, or even required, but balance confidentiality remains essential. Moreover, they enable optional auditable encryption: each transferred amount can be encrypted not only for the recipient but also for a designated auditor using a dedicated public key. This optional regulatory channel allows a trusted entity to decrypt values when required, for compliance, enforcement, or dispute resolution, without compromising broader privacy. 

To achieve this, we use ElGamal over elliptic curves (chosen for its additive homomorphic property) to conceal transaction amounts, while leveraging zero-knowledge proofs to ensure correctness and enforce balance integrity. This allows the sender to update their own and the recipient's balances without revealing the actual transferred amount, while proving that it is non-negative, within range, and that the encrypted update of the remaining balance is correctly computed (i.e., subtraction and addition match the same plaintext amount).

The protocol binds all operations to on-chain accounts, ensuring that private asset control is always linked to verifiable, authorized actions. Taken together, Haults serve as a viable and lightweight foundation for confidential assets in permissioned smart contract environments, bridging the gap between privacy and compliance. 

\subsection{Roadmap}

The document is organized as follows. In Section~\ref{sec:buildingblocks}, we describe the main elements and cryptographic primitives used by Haults. In Section~\ref{sec:solution}, we describe our solution in detail. In Section~\ref{sec:security}, we provide an informal security discussion. In Section~\ref{sec:deposits}, we extend the features of our solution. In Section~\ref{sec:implementation}, we describe the main implementation details concerning our solution's proof-of-concept, as well as some results. We conclude in Section~\ref{sec:conclusions}.

\section{Building Blocks}
\label{sec:buildingblocks}
In this section, we describe the main building blocks and cryptographic primitives used by Haults. In Section~\ref{sec:concrete:ecc}, we define the elliptic curve notation used throughout the document. In Section~\ref{sec:concrete:encryption}, we describe the homomorphic encryption scheme used to encrypt balances. Finally, in Section~\ref{sec:solution:maps}, we define the mapping functions used to convert scalar values into elliptic curve points, which are required for encryption.

\subsection{Elliptic Curves} \label{sec:concrete:ecc}

Our solution's security relies on the security of elliptic curves on different levels. While they are used on the Blockchain layer to compute and verify zero-knowledge proofs, and to do elliptic curve operations in-circuit, we will set their notation now for the sake of solely understanding our main contribution.

We call $E$ an elliptic curve over the finite field $\mathbb{F}_{q}$, where $q$ is a large prime number, and denote this by $E/\mathbb{F}_{q}$. With this setting, we call $\G$ its prime subgroup of large prime order $r$, and $G$ the generator of such a group.

We also define a function to verify whether a point $P = (x,y)$ belongs to the curve $E$, and in particular, to the desired subgroup. It is defined as follows:

\begin{definition}[Check elliptic curve point validity] Let $E/\mathbb{F}_{q}$ be an elliptic curve with a subgroup $\G$ of order $r$, and let $P = (x, y)$ be an arbitrary point. We define a function returning 1 if $P$ is a valid point belonging to the desired subgroup, and 0 otherwise, as follows:
	\begin{description}
		\item $0/1 \leftarrow \CheckPoint(P)$: 
		\begin{itemize}
			\item If $P = \mathcal{O}$, where $\mathcal{O}$ is the point at infinity, return 0.
			\item If $P$ does not satisfy the curve equation of $E$, so that it does not belong to the curve, return 0.
			\item If $r \cdot P = \mathcal{O}$, return 1 (as $P$ is in the subgroup of order $r$). Otherwise, return 0.
		\end{itemize}
	\end{description}
\end{definition}

\subsection{Homomorphic Encryption} \label{sec:concrete:encryption}

Haults use ElGamal~\cite{elgamal1985public} over elliptic curves (EC) to encrypt balances. We first define the functions we use to encrypt and decrypt messages. 

\begin{definition}[EC-ElGamal encryption] Let $\sk \sample \F_r$ be a secret key and  \(\pk = \sk \cdot G \in \mathbb{G}\) be a public key. Given \(M \in \mathbb{G}\) being a message curve point, the encryption function $\Enc$ and the decryption $\Dec$ are defined as follows.
	\begin{description}
		\item $(C_1, C_2) \leftarrow \Enc_{\pk}(M \in \G)$: 
		\begin{enumerate}
			\item Sample a random scalar $k \sample \F_r$.
			\item Compute $C_1 = k \cdot G$.
			\item Compute $C_2 = k\cdot \pk + M$. 
			\item Output $(C_1, C_2)$.
		\end{enumerate}

		\medskip
		\item $M \leftarrow \Dec_{\sk}((C_1, C_2))$: 
		\begin{enumerate}
			\item Compute $K = \sk \cdot C_1$.
			\item Compute $M = C_2 - K$.
			\item Return $M$.
		\end{enumerate}
	\end{description}
\end{definition}

Homomorphic encryption is the backbone of our protocol. In particular, we use the additive homomorphic property of EC-ElGamal, demonstrated as follows. Let \(M_1,M_2 \in \mathbb{G}\) be two messages with the following encryptions: 

	$$\Enc_{\pk}(M_1) = (C_{1,1}, C_{1,2})$$
	$$\Enc_{\pk}(M_2) = (C_{2,1}, C_{2,2})$$

Given the ciphertexts component-wise additions, we have that:

	$$\Enc_{\pk}(M_1) + \Enc_{\pk}(M_2) = (C_{1,1} + C_{2,1}, C_{1,2} + C_{2,2}) = \Enc_{\pk}(M_1 + M_2)$$

and thus, EC-ElGamal is additively homomorphic on the elliptic curve group. To prove that, we take the initial elements:

	$$\Enc_{\pk}(M_1) = (C_{1,1}, C_{1,2}) = (k_1 \cdot G, k\cdot \pk + M_1)$$
	$$\Enc_{\pk}(M_2) = (C_{2,1}, C_{2,2}) = (k_2 \cdot G, k\cdot \pk + M_2)$$

and compute the addition:

	$$(C_{1,1} + C_{2,1}, C_{1,2} + C_{2,2}) = (k_1G + k_2G, M_1 + k_1 \cdot \pk + M_2 + k_2 \cdot \pk) =$$
	$$= ((k_1 + k_2)G, M_1 + M_2 + (k_1 + k_2) \cdot \pk)$$
	
which is the encryption of $M_1 + M_2$ using randomness $k = k_1 + k_2$. As such, addition of EC-ElGamal ciphertexts yields an encryption whose original plaintext is the sum of the original plaintexts.

\subsection{Mappings} \label{sec:solution:maps}

Since ElGamal requires the input to be a curve point, we define two mapping functions from scalar values to elliptic curve points. The first variant $\MapHomomorphic$ will allow homomorphic addition of ciphertexts but not direct recovery of the plaintext value.

\begin{definition}[Homomorphic mapping] Let $v \in \F_r$ be an input value to map. We define an homomorphic mapping function as follows:
	\begin{description}
		\item $M \leftarrow \MapHomomorphic(v)$: 
		\begin{enumerate}
			\item Map $v$ to $\G$ via scalar multiplication: $M = v \cdot G$.
			\item Return $M$.
		\end{enumerate}
	\end{description}
\end{definition}

The second variant $\MapRecoverable$ will enable the recipient or auditor to recover the original value \(v\) by inspecting the least significant $b$ bits of the \(y\)-coordinate of \(M\). For convenience, we first define the function  $\LSB_b$, which does exactly this. Let $M \in \G$ be a mapped value, from which we want to recover the original value of maximum size $b$-bits.
	
\begin{definition}[Least significant bits extraction] We define a function to extract the least significant $b$ bits from the $y$-coordinate of an elliptic curve point \(M\) representing a mapped value:
	\begin{description}
		\item $v \leftarrow \LSB_{b}(M)$: 
		\begin{enumerate}
			\item Return the least significant $b$-bits of the $y$-coordinate of $M$.
		\end{enumerate}
	\end{description}
\end{definition}

\begin{definition}[Recoverable mapping] Let $v_b \in \F_r$ be an input value to map, which maximum bit size is $b$ bits, i.e., $v_b \in [0, 2^b)$. We define a recoverable mapping function as follows:
	\begin{description}
		\item $M \leftarrow \MapRecoverable(v_b)$: 
		\begin{enumerate}
			\item Find $M$ in $\G$ such that $\LSB_b(M) = v$.
			\item Return $M$.
		\end{enumerate}
	\end{description}
\end{definition}

\section{Our Solution: Haults} 
\label{sec:solution}
In this section, we describe the Hault protocol in detail. In Section~\ref{sec:solution:participants}, we describe all three roles that participants of our protocol can take. In Section~\ref{sec:solution:keys}, we define the involved cryptographic keys and how they are managed by the contract. In Section~\ref{sec:solution:notes}, we present the structure of Hault notes, which encapsulate encrypted units of value and form the core mechanism for private balance management and transfers. In Section~\ref{sec:solution:transfer-flow}, we describe the transfer workflow. In Section~\ref{sec:solution:circuit}, we describe the zero-knowledge circuit used to enforce the correctness and integrity of encrypted transactions. Finally, the \textit{minting} and \textit{force} transfers are described in Sections~\ref{sec:solution:mint} and~\ref{sec:solution:force}, respectively.

\subsection{Participants} \label{sec:solution:participants}

There are three types of participants in a Hault, the \textit{users}, the \textit{contract owner}, and the \textit{auditor}:

\begin{itemize}
	\item \textbf{Users:} they can hold notes representing their encrypted balance, and can perform transfers to other users within the Hault. Users can only interact with the Hault if they are part of the allowlist managed by the contract owner.
	\item \textbf{Contract owner:} a participant who has the authority to add users to the allowlist and to remove them. It also takes the role of an \textit{issuer}, being the only participant allowed to mint new notes within the Hault, effectively increasing the total supply.
	\item \textbf{Auditor:} a participant responsible for regulatory oversight within the Hault. All transfers include an encryption of the transferred amount for the auditor, allowing them to decrypt and review transactions when necessary. Plus, it is the only participant allowed to perform \textit{force} transfers, which enable the auditor to move funds between users with or without their consent, to recover the funds from an account with a lost secret key, or for compliance purposes.
\end{itemize}

\subsection{Keys and Identity Management} \label{sec:solution:keys}
In addition to the target Blockchain native keypair \((\sk^\nat, \pk^\nat)\) over a given curve, each Hault user also maintains a Hault-specific keypair used for encryption. The Hault keypair is defined as:
\[
(\sk, \pk) \in \mathbb{F}_r \times \mathbb{G} \quad \text{where} \quad \pk = \sk \cdot G
\]
Here, \(\sk\) is a private scalar in the field \(\mathbb{F}_r\), and \(\pk\) is the corresponding public key on the elliptic curve group \(\mathbb{G}\).
The native public key \(\pk^\nat\) serves as a pointer to the user’s Hault identity. To authorize any Hault transaction, users must sign transaction with their private native key \(\sk^\nat\).

Similarly, the auditor $D$ also holds a Hault keypair \((\sk[D], \pk[D]) \in \mathbb{F}_r \times \mathbb{G}\), where \(\pk[D] = \sk[D] \cdot G\). This key is used to decrypt auditor-encrypted values and perform regulatory functions within the Hault.

\paragraph{Adding a user to the allowlist.} To add a user to the allowlist, the contract owner calls a contract method (only callable by the owner), providing the user’s native public key \(\pk^\nat\) and their Hault public key \(\pk\). Then contract performs the following steps:

\begin{itemize}
	\item Checks that the \(\pk^\nat\) is not already in the allowlist.
	\item Adds a map \(\pk^\nat\) to \(\pk\) to the allowlist, along with a flag enabling the user to interact with the Hault.
\end{itemize}

Ideally, the contract should also check if \(\CheckPoint(\pk)\) returns true. However, this includes a scalar multiplication by the curve order $r$, which is too expensive to be done on-chain. Thus, this task gets delegated off-chain to the contract owner, who must ensure that the provided public keys are valid before adding them to the allowlist. In the case that the assumption that all the checks are correctly performed by the owner does not hold, this check could also be enforced within the zero-knowledge transfer circuit (see Section~\ref{sec:solution:circuit}).

\subsection{Notes} \label{sec:solution:notes}

Within Haults, balances are represented and transferred via data structures called \textit{notes}. Each note owned by a user encapsulates a given value $v$ in the form of encrypted data. In particular, a two different ciphertexts:

\[
\note = (v_{\enc}, V_{\enc}),
\]
where
\begin{itemize}
	\item \(v_{\enc}\) is an encryption of the value using the map $\MapRecoverable$, enabling the owner to decrypt it and to retrieve the plaintext.
	\item \(V_{\enc}\) is an encryption of the same value using the map $\MapHomomorphic$, allowing homomorphic addition without prior decryption.
\end{itemize}

\paragraph{Balances.} Now, let us set a user with Hault keys \((\sk, \pk)\). The user’s total encrypted balance is defined as the sum of all the homomorphic encryptions \(V_{\enc}\) associated with the notes they own:
\[
V_{\enc} = \sum_i \note_i.V_{\enc}
\]
To obtain the plaintext balance, the user decrypts the encryptions $v_{\enc}$ with their secret key to obtain the result:
\[
v = \sum_{i} \LSB_b(\Dec_{\sk}(\note_i.v_{\enc})) \quad \text{such that} \quad v \cdot G = \Dec_{\sk}(V_{\enc}).
\]

\paragraph{Auditor encryption.} In every transaction, the transferred value is also encrypted using the auditor's public key \(\pk[D]\). This ciphertext is included as:
\[
(v_{\enc[D]}, V_{\enc[D]})
\]
where both components are constructed in the same manner as for regular recipients, but using \(\pk[D]\) as the encryption key. This enables a designated auditor $D$ to retrieve transfer values when necessary, e.g., for regulatory compliance.

\subsection{Transfers} \label{sec:solution:transfer-flow}

In this section, we describe the process by which a user (the sender) transfers a value $w$ to another user (the recipient) within the Hault. The transfer involves consuming previously unspent notes, generating new ones, and producing a zero-knowledge proof to ensure correctness, integrity, and compliance. It is described in Protocol~\ref{prot:transfer}.

\begin{protocol}[Hault transfer]
\label{prot:transfer}
Let \(A\) be the sender holding the Hault keys $(\sk[A], \pk[A])$ and \(B\) the recipient with keys $(\sk[B], \pk[B])$. The sender \(A\) holds a set of $n$ notes
\[
\left\{\note_i = (v_{\enc[A]},  V_{\enc[A]})\right\}_{i=1}^{n}.
\]

The sender $A$ will create a new note for \(B\) with value \(w\), and all their notes will be removed and replaced by a new note owned by \(A\) that holds A's current balance minus \(w\). To issue a transfer, the sender \(A\) performs the following steps:

\begin{enumerate}
	\item \textbf{Map $w$ to points in the curve.}
		\begin{itemize}
			\item Compute $w_{\recov} = \MapRecoverable(w)$.
			\item Compute $w_{\homom} = \MapHomomorphic(w)$.
		\end{itemize}	
	\item \textbf{Aggregate sender notes.} 
	\begin{itemize}
		\item Compute the total \textit{old} value held in the unspent notes:
		$$
		v^{\old} = \sum_{i} \LSB(\Dec_{\sk[A]}(\note_i.v_{\enc[A]}))
		\hspace{0.5em} \text{and} \hspace{0.5em}
		V^{\old}_{\enc[A]} = \sum_i \note_i.V_{\enc[A]}.
		$$
		\item Compute a new note containing the sender’s updated balance, by subtracting the transfer amount \(w\), and yielding the new encrypted balance:			
		$$
		v_{\enc[A]}^{\new} = \Enc_{\pk[A]}(\MapRecoverable(v^{\old} - w))
		\quad \text{and} \quad
		V_{\enc[A]}^{\new} = V^{\old}_{\enc[A]} - \Enc_{\pk[A]}(w_{\homom}).
		$$
	\end{itemize}
	\item \textbf{Create a new note for the recipient.}
	\begin{itemize}
		\item Compute a new note for the recipient \(B\), encrypted using \(\pk[B]\):
		$$
		w_{\enc[B]} = \Enc_{\pk[B]}(w_{\recov})
		\quad \text{and} \quad
		W_{\enc[B]} = \Enc_{\pk[B]}(w_{\homom}).
		$$
		\item Compute a copy of the same note encrypted for the auditor using \(\pk[D]\):
		$$
		w_{\enc[D]} = \Enc_{\pk[D]}(w_{\recov})
		\quad \text{and} \quad
		W_{\enc[D]} = \Enc_{\pk[D]}(w_{\homom}).
		$$
	\end{itemize}
	\item \textbf{Generate a zero-knowledge proof $\pi$.} Use the circuit from Section~\ref{sec:solution:circuit} to prove that all the operations were correct (i.e. that the transfer is valid) without revealing the value $w$.
	\item \textbf{Call the transfer method.} Send the zero-knowledge proof $\pi$ along with the public inputs $\PI$ (that contain old the encryptions and the public keys) to the Hault smart contract. The smart contract performs the following operations.
	
	\begin{itemize}
		\item Verify that the sender and recipient native addresses are allowed to transact, and that their Hault addresses are $\PI.\pk[A]$ and $\PI.\pk[B]$ respectively.
		\item Verify that the auditor's public key is $\PI.\pk[D]$. 
		\item Check that 
		$$\PI.V^{\old}_{\enc[A]} \stackrel{?}{=} \sum_i \note_i.V_{\enc[A]}.$$
		\item Verify the zero-knowledge proof $\pi$.
		\item Remove all old notes $\{\note_i\}_{i=1}^{n}$ owned by $A$ from their state, and append the new note
		$$\note = (\PI.v_{\enc[A]}^{\new}, \PI.V_{\enc[A]}^{\new}).$$
		\item Remove all old notes $\{\note_i\}_{i=1}^{n}$ owned by $A$ from their auditable state, and append the new note
		$$\note = (\PI.v_{\enc[D]}^{\new}, \PI.V_{\enc[D]}^{\new}).$$
		\item Append to the recipient's state the new note
		$$\note = (\PI.w_{\enc[B]}, \PI.W_{\enc[B]}).$$
		\item Append to the recipient's auditable state the new note 
		$$\note = (\PI.w_{\enc[D]}, \PI.W_{\enc[D]}).$$
	\end{itemize}
\end{enumerate}
\end{protocol}

\subsection{Transfer Circuit} \label{sec:solution:circuit}

\medskip

We describe the circuit that uses the sender to prove in zero-knowledge that their transaction is correct. We start by defining the circuit constants, then the public and private inputs, and finally the constraints that need to be satisfied.

\medskip

\noindent
\textbf{Constants}
\begin{itemize}
	\item {\color{orange!80!black}{$G$}}: generator of $\G$ as defined in Section~\ref{sec:concrete:ecc}. 
	\item {\color{orange!80!black}{$b$}}: maximum bit-length for values.
\end{itemize}
Note that the generator is used in the circuit to perform different operations such as encryption, but we omit it in the constraints explanation for clarity. 

\medskip
\noindent
\textbf{Public inputs}
\begin{itemize}
	\item {{\color{blue!80!black}{$\pk[A], \pk[B], \pk[D]$}}}: public keys of the sender, recipient, and auditor, respectively. 
	\item {{\color{blue!80!black}{$v_{\enc[A]}^{\new}, V_{\enc[A]}^{\new}$}}}: encryption of the new balance of the sender.
	\item {{\color{blue!80!black}{$w_{\enc[B]}, W_{{\enc[B]}}$}}}: encryption of $w$ for the recipient.
	\item {{\color{blue!80!black}{$w_{\enc[D]}, W_{\enc[D]}$}}}: encryption of $w$ for the auditor.
	\item {{\color{blue!80!black}{$V_{\enc[A]}^{\old}$}}}: encryption of the old balance of the sender.
\end{itemize}
\textbf{Private inputs}
\begin{itemize}								
	\item $\sk[A]$: secret key of the sender.
	\item $w_{\recov}$: point whose last $b$ bits of its $y$-coordinate corresponds to the value $w$.
	\item $v_{\recov}^{\new}$: point whose last $b$ bits of its $y$-coordinate is the old balance of the sender.
	\item $v^{\old}$: the old balance of the sender.
\end{itemize}
Note that the user will have to provide the randomness used in the encryptions as private inputs as well, but we omit them here for clarity.

\medskip

\noindent
\textbf{Constraints}
\begin{itemize}	
	\item Creating a new note
	\begin{itemize}	
		\item Check that $\CheckPoint(w_{\recov}) \stackrel{?}{=} 1$.
		\item Let $w = \LSB_{\color{orange!80!black}{b}}(w_{\recov})$, while enforcing $w \in [0,2^{\color{orange!80!black}{b}})$.	
		\item Compute $w_{\homom} = \MapHomomorphic(w)$.
		\item Compute ${\color{blue!80!black}{w_{\enc[B]}}} = \Enc_{{\color{blue!80!black}{\pk[B]}}}(w_{\recov})$ and ${\color{blue!80!black}{W_{\enc[B]}}} = \Enc_{\color{blue!80!black}{\pk[B]}}(w_{\homom}).$
	\end{itemize}
	
	\item Aggregating old notes
	\begin{itemize}	
		\item Ensure $v^{\old} \in [0,2^{\color{orange!80!black}{b}})$
		\item Compute ${\color{blue!80!black}{\pk[A]}} = \sk[A] \cdot {\color{orange!80!black}{G}}$.
		\item Check that $\MapHomomorphic(v^{\old}) \stackrel{?}{=} \Dec_{\sk[A]}({\color{blue!80!black}{V_{\enc[A]}^{\old}}})$.
	\end{itemize}

	\item Computing new balance
	\begin{itemize}	
		\item Compute $v^{\new} = v^{\old} - w$.
		\item Check that $\CheckPoint(v_{\recov}^{\new}) \stackrel{?}{=} 1$.
		\item Check that $v^{\new} \stackrel{?}{=} \LSB_{\color{orange!80!black}{b}}(v_{\recov}^{\new})$, while enforcing $v^{\new} \in [0,2^{\color{orange!80!black}{b}})$.
		\item Compute $W_{\enc[A]} = \Enc_{\color{blue!80!black}{\pk[A]}}(w_{\homom})$
		\item Compute ${\color{blue!80!black}{V_{\enc[A]}^{\new}}} = {\color{blue!80!black}{V_{\enc[A]}^{\old}}} - W_{\enc[A]}$. This operation is not really needed in the circuit, but it allows us to save a subtraction in the contract.
		\item Compute ${\color{blue!80!black}{v_{\enc[A]}^{\new}}} = \Enc_{\color{blue!80!black}{\pk[A]}}(v_{\recov}^{\new}).$
	\end{itemize}

	\item Encrypting for the auditor
	\begin{itemize}	
		\item Compute ${\color{blue!80!black}{v_{\enc[D]}^{\new}}} = \Enc_{\color{blue!80!black}{\pk[D]}}(v_{\recov}^{\new})$ and ${\color{blue!80!black}{V_{\enc[D]}^{\new}}} = \Enc_{\color{blue!80!black}{\pk[D]}}(\MapHomomorphic(v^{\new})).$
		\item Compute ${\color{blue!80!black}{w_{\enc[D]}}} = \Enc_{\color{blue!80!black}{\pk[D]}}(w_{\recov})$ and ${\color{blue!80!black}{W_{\enc[D]}}} = \Enc_{\color{blue!80!black}{\pk[D]}}(w_{\homom}).$
	\end{itemize}
\end{itemize}

\subsection{Minting New Notes} \label{sec:solution:mint}

The minting process is performed by the contract owner, who takes the role of an issuer. The process to mint new notes is similar to a transfer, with the main difference that there is no sender consuming old notes. Instead, the contract owner creates a new note with a given value \(w\) for a recipient \(B\). To avoid requiring two different circuits to maintain, we define the minting process in a way that reuses the normal transfer circuit while serving the minting purpose. In order to allow anyone to verify the token supply, we introduce the concept of \textit{transparently encrypted notes}: they leak the note amount, while keeping the same structure as other notes have, so they can be aggregated and hidden into normaly encrypted notes. The minting of a new note with value $w$ for a given recipient works in a similar way to a transfer (as described in Protocol~\ref{prot:transfer}), with the following differences:

\begin{itemize}
	\item We set an old balance so that it simply verifies, i.e., $v^{\old} \ge w$, and compute the related $V_{\enc[A]}^{\old}$.
	\item We compute the zero-knowledge proof using the transfer circuit from Section~\ref{sec:solution:circuit}, with the sender's public key $\pk[A]$ corresponding to the contract owner's key. However, we now use a randomness $r = 0$ to compute the transparent encryption $w_{\enc[B]}$.
	\item We call the minting method, and send the zero-knowledge proof $\pi$ along with the public inputs $\PI$ to the Hault smart contract. The smart contract performs exactly the following operations.
	\begin{itemize}
		\item Verify that the sender native address corresponds to the contract owner, and that their Hault address is $\PI.\pk[A]$.
		\item Verify that the recipient's native address is allowed to transact, and that their Hault address is $\PI.\pk[B]$.
		\item Verify that the auditor's public key is $\PI.\pk[D]$. 
		\item Check that $\PI.w_{\enc[B]}.C_1 \stackrel{?}{=} \mathcal{O}$ (identity point).
		\item Verify $\pi$.
		\item Append the new $\note = (\PI.w_{\enc[B]}, \PI.W_{\enc[B]})$ to the recipient's state.
		\item Append the new $\note = (\PI.w_{\enc[D]}, \PI.W_{\enc[D]})$ to the recipient's auditable state.
		\item Update the total supply of the contract by adding $w = \LSB_b(\PI.w_{\enc[B]}.C_2)$.
	\end{itemize}	
\end{itemize}

\subsection{Force Transfers} \label{sec:solution:force}

The force transfer process is performed by the auditor, who has the authority to move funds between users, either with or without their consent. This mechanism is crucial for recovering funds from accounts with lost secret keys or for compliance purposes. It basically performs a transfer by using the target's auditable balances, while proving the same consistency checks that a regular transfer does. For simplicity, and to avoid mantaining two different circuits, the force transfer can reuse the transfer circuit to completely move all the funds from a user \(A\) to another user \(B\). It is done by calling a force transfer method in the contract, that performs a regular transfer, with the following differences:

\begin{itemize}
	\item The sender is the auditor, thus the sender's keypair $\sk[A], \pk[A]$ corresponds to the auditor's keys $\sk[D], \pk[D]$.
	\item The old balance is retrieved from the auditable state of user \(A\), i.e. $V_{\enc[D]}^{\old}$ .
	\item All notes are removed from the state of user \(A\), and no new balance is appended.
	\item The auditor uses a randomness $r = 0$ to compute the encryption $v_{\enc[D]}^{\new}$. And then the contract checks both that 
	
	$$v_{\enc[D]}^{\new}.C_1 \stackrel{?}{=} \mathcal{O} \quad \text{and} \quad \LSB_b(v_{\enc[D]}^{\new}.C_2) \stackrel{?}{=} 0.$$
\end{itemize}

\section{Security Discussion}
\label{sec:security}

In this section, we informally outline and discuss the main security goals of the Hault protocol. We start by defining our threat model, and then we discuss how the protocol achieves the desired security properties.

\paragraph{Threat model.}
We assume a Hault where anyone can read all the issued transactions. We assume that only allowed users, those who have been added to the allowlist by the contract owner, can call contract methods and interact with the protocol. We assume all adversaries to be probabilistic polynomial time (PPT). Plus, we consider all users' secret keys to be kept private and secure.

Adversaries who are not part of the allowlist can try to read hidden user balances or transferred amounts. Adversaries who are part of the allowlist can attempt to create invalid transactions and try to read hidden user balances or transferred amounts.

All the cryptographic primitives used by the protocol are assumed to be secure: the involved elliptic curves are secure under the DLOG assumption, the ElGamal encryption scheme is IND-CPA secure, and the selected zero-knowledge proof system is complete, sound, and zero-knowledge.

\paragraph{Completeness.}
An honest user who generates a transaction according to the protocol will always have the transaction accepted by the network.

\paragraph{Soundness.}
No adversary can construct a transaction that passes verification while spending others' balance, unless we contradict the assumption that the secret keys are kept private and secure. This is enforced on two levels:
\begin{itemize}
    \item the contract enforces that the user triggering the transaction (thus knowing the secret key) can only spend notes from their own state;
    \item the zero-knowledge proof enforces the user's own public key to be computed from their Hault secret key.
\end{itemize} 

On the other hand, no adversary can construct a transaction that passes verification while creating money out of thin air. This is enforced by the homomorphic addition of old notes, whose decryption binds to a deterministic total amount.

\paragraph{Balance integrity.}
The integrity of the balances cannot be tempered, since the zero-knowledge proof enforces that for a user's decrypted old balance $v^{\old}$, the new balance $v^{\new}$ is computed by subtracting the transferred amount $w$, while being kept within the allowed range $[0, 2^b)$.

\paragraph{Double-spending resistance.}
Notes get removed from the state once spent, and the contract enforces that only the remaining ones can be used. Given the sequential nature of transaction processing by the Blockchain, an adversary cannot reuse notes that have already been spent in previous transactions.

\paragraph{Zero-knowledge and readability.}
The old balance $v^{\old}$ and the new balance $v^{\new}$ remain hidden to everyone except the user owning the balance. The transferred amount $w$ remains hidden to everyone except the sender and the recipient. This is enforced by the zero-knowledge property of the zero-knowledge proof system, which guarantees that no information about these values is leaked during the verification of the transaction. For what concerns readability, only a given note's owner can decrypt the value $v$ from its corresponding ciphertext, as they are the only one knowing the corresponding secret key.

\paragraph{Non-malleability.}
While ElGamal ciphertexts are malleable by nature, the zero-knowledge proof binds the original computed ciphertexts to the proof. Moreover, the zero-knowledge proof binds to a given contract state and to the user owning the secret key. This prevents an adversary from computing proofs using different ciphertexts or notes than those included in the transaction, and from performing replay attacks using intercepted proofs.

\section{Deposits and Withdrawals}
\label{sec:deposits}
In this section, we extend our solution to support deposits and withdrawals of native assets (e.g., Ether or ERC-20 tokens), so that they can be moved into the Hault to hide them, and send them to other parties privately. However, it is important to notice that withdrawals can partially or totally break the zero-knowledge property of the protocol, as the withdrawn amount becomes public, thus leaking information not solely about what the user executing the withdrawal but about the entire system state. To overcome this problem, we envisage this property to be used in the context of real-world assets (RWA), where a trusted entity (the contract owner) is in charge of managing deposits and withdrawals of off-chain assets and to distribute them to on-chain users.

\paragraph{Deposits.}
The process to deposit native VM tokens into the Hault is similar to the minting process described in Section~\ref{sec:solution:mint}, where the contract owner deposits native assets into the Hault contract by creating new temporary transparent notes (that get hidden once sent to another party). Here, the contract owner follows the exact same steps, with the difference that includes a payment into the transaction. In other words, the called method is payable, and the amount of native assets sent with the transaction is the one that gets encrypted into the created note. As such, beyond proof verification and security checks, the contract also verifies that the amount of native assets sent with the transaction matches the transparently encrypted amount in the created note. The same approach can be extended to support other tokens, such as ERC-20.

\paragraph{Withdrawals.}
To withdraw native VM tokens from the Hault, like happens with the deposit, the contract owner follows the exact same steps that a mint requires, with the following differences: this time, the old balance has to be consistent, like happens with transfers, and once the proof is verified, the contract sends the transparently encrypted amount to a given recipient address. Again, the same approach can be extended to support other tokens, such as ERC-20.

\section{Implementation and Benchmarks}
\label{sec:implementation}
In this section, we describe the main considerations to take into account when implementing the Hault protocol, as well as some benchmarks obtained from our proof-of-concept.

\medskip
We have implemented a proof-of-concept of the Hault protocol using Solidity for the smart contract, targeting EVM-enabled Blockchains, and Circom~\cite{10002421} for the zero-knowledge proof circuit. The zero-knowledge proof scheme is Groth16~\cite{cryptoeprint:2016/260}, and the chosen setting of elliptic curves is the BN254~\cite{cryptoeprint:2005/133} for the proof, and Baby JubJub~\cite{bjubjub} for elliptic curve operations in-circuit. BN254 is natively supported by the EVM precomplies, and by the implementation of the prover we used, snarkjs\footnote{https://github.com/iden3/snarkjs}. Nonetheless, the protocol is not tied to these specific technologies and can be adapted to other environments and proving systems. It is worth noticing that, since the Pectra upgrade\footnote{https://ethereum.org/roadmap/pectra/} on May 7th, 2025, Ethereum and most EVMs include BLS12-381 precomplies. As such, the pair BLS12-381\footnote{https://github.com/zkcrypto/bls12\_381} / JubJub\footnote{https://github.com/zkcrypto/jubjub}, which provides better security levels, would be a better choice in production.

\paragraph{Benchmarks.}
The benchmarks we conducted focus on the gas spent by the contract methods, and the time it takes to generate the zero-knowledge proofs. For what concerns the circuit, we implemented it using \textbf{less than $2^{16}$ constraints}. The proving time for computing a transfer proof using snarkjs in Firefox, using an \textit{AMD Ryzen 7 5800X}, is under \textbf{$\sim$2 seconds on average}, making our protocol compatible with end-user devices.

The gas spent by the contract to verify the proof is \textbf{$\sim$503,948 gas}, and a complete transfer \textbf{$\sim$894,502}. While this cost is way higher than non-private transfers, it is still affordable in many scenarios. Moreover, some solutions to reduce gas costs include using layer-2 Blockchains, or deploying the proof verifier as a cheaper precompile, if this is an option in the target Blockchain.

\section{Conclusions}
\label{sec:conclusions}
In this work, we presented a protocol that enables anonymous balances and transaction amounts. We demonstrated its efficiency through several benchmarks, as well as describing its security properties in detail. As we have seen, deposits and withdrawals of external tokens are compatible with Haults, with the caveat that withdrawals may leak some information about the system state, and thus should be handled by contract owners. Therefore, further research is needed to design a secure and private mechanism for deposits and withdrawals in other scenarios as well. Finally, our protocol can also serve as a building block for privacy-preserving financial applications, like confidential order books or any other decentralized application involving money transactions. Plus, it can be used as a foundation for further research in the field of on-chain protocols, where computation on hidden values is a requirement.

\section*{Acknowledgements}
This work has been partially funded by Dusk Network B.V.. We also thank Marta Bellés-Muñoz for her feedback on this work.

%
\bibliographystyle{splncs04}
\bibliography{bibtex}

\end{document}